\documentclass[12pt]{article}%
\usepackage[nosort]{cite}
\usepackage{graphicx}
\usepackage{multicol}
\usepackage{amsfonts}
\usepackage{amssymb}
\usepackage{amsmath}
\usepackage{torres}
\usepackage{slashed}
\usepackage{setspace}
\usepackage{verbatim}
\usepackage{color}
\usepackage{longtable}
\usepackage{float}
\usepackage{braket}
\usepackage{subcaption}
\usepackage{epsfig}
\usepackage{epstopdf}
\usepackage{adjustbox}
\usepackage{tikz}
\usepackage[margin=1in]{geometry}
\usepackage{titletoc}
\usepackage{hyperref}%
\usepackage{mathrsfs}
\usepackage{dsfont}
\usepackage{algorithm}
\usepackage{algpseudocode}
\usepackage{upgreek}
\usepackage{mathtools}
\usepackage{stmaryrd}
\usepackage{pdflscape}

\pdfoutput=1

\usepackage{caption}
\providecommand{\U}[1]{\protect\rule{.1in}{.1in}}

\newsavebox{\mysavebox}

\hypersetup{colorlinks,citecolor=black,filecolor=black,linkcolor=black,urlcolor=black}
\usetikzlibrary{decorations.markings}
\numberwithin{equation}{section}

\usepackage{tikz}
\usetikzlibrary{arrows}
\usetikzlibrary{arrows.meta}
\usetikzlibrary{arrows,decorations.pathmorphing}
\usetikzlibrary{shapes.geometric,calc,arrows, positioning,shapes.misc,decorations.markings}
\tikzset{
  big arrow/.style={
    decoration={markings,mark=at position 1 with {\arrow[scale=2,#1]{>}}},
    postaction={decorate},
    shorten >=0.4pt},
  big arrow/.default=black}

\usetikzlibrary{hobby}
\usetikzlibrary{decorations.markings}

\pgfdeclarelayer{edgelayer}
\pgfdeclarelayer{nodelayer}
\pgfsetlayers{edgelayer,nodelayer,main}
\tikzstyle{none}=[inner sep=0pt]

\tikzstyle{NodeCross}=[draw, shape=circle, cross out, inner sep=0pt, minimum size=5pt,line width=0.25mm]
\tikzstyle{SmallCircle}=[draw, shape=circle, black, fill=black, inner sep=0pt, minimum size=6pt]
\tikzstyle{BigCircle}=[draw, shape=circle, black, fill=black, inner sep=0pt, minimum size=20pt]
\tikzstyle{SmallCircleRed}=[draw, shape=circle, fill={rgb,255: red,191; green,0; blue,0}, inner sep=0pt, minimum size=6pt]
\tikzstyle{BigCircleRed}=[draw, shape=circle, fill={rgb,255: red,191; green,0; blue,0}, inner sep=0pt, minimum size=10pt]
\tikzstyle{SmallCircleBlue}=[draw, shape=circle, fill=blue, inner sep=0pt, minimum size=6pt]
\tikzstyle{BigCircleBlue}=[draw, shape=circle, fill=blue, inner sep=0pt, minimum size=10pt]
\tikzstyle{SmallCirclePurple}=[draw, shape=circle, fill={rgb,255: red,191; green,0; blue,191}, inner sep=0pt, minimum size=6pt]
\tikzstyle{BigCirclePurple}=[draw, shape=circle, fill={rgb,255: red,191; green,0; blue,191}, inner sep=0pt, minimum size=10pt]
\tikzstyle{SmallCircleGreen}=[draw, shape=circle,  fill={rgb,255: red,80; green,200; blue,120}, inner sep=0pt, minimum size=6pt]
\tikzstyle{BigCircleGreen}=[draw, shape=circle,  fill={rgb,255: red,80; green,200; blue,120}, inner sep=0pt, minimum size=10pt]
\tikzstyle{SmallCircleBrown}=[draw, shape=circle,  fill={rgb,255: red,210; green,105; blue,30}, inner sep=0pt, minimum size=6pt]
\tikzstyle{BigCircleBrown}=[draw, shape=circle,  fill={rgb,255: red,210; green,105; blue,30}, inner sep=0pt, minimum size=8pt]
\tikzstyle{Star}=[draw, shape=star, fill=black, star points=8, inner sep=0pt, minimum size=10pt]

\tikzstyle{DashedLine}=[-, densely dashed, line width=0.25mm]
\tikzstyle{DottedLine}=[-, dotted, line width=0.25mm]
\tikzstyle{ThickLine}=[-, line width=0.25mm]
\tikzstyle{RedLine}=[-, draw={rgb,255: red,191; green,0; blue,0}, fill=none, line width=0.5mm]
\tikzstyle{DashedRedLine}=[-, densely dashed, draw={rgb,255: red,191; green,0; blue,0}, fill=none, line width=0.5mm]
\tikzstyle{DottedRed}=[-, dotted, draw={rgb,255: red,191; green,0; blue,0}, fill=none, dotted, line width=0.5mm]
\tikzstyle{BlueLine}=[-, draw=blue, fill=none, line width=0.5mm]
\tikzstyle{DashedBlueLine}=[-, densely dashed, draw=blue, fill=none, line width=0.5mm]
\tikzstyle{DottedBlue}=[-, dotted, draw=blue, fill=none, dotted, line width=0.5mm]
\tikzstyle{PurpleLine}=[-, draw={rgb,255: red,191; green,0; blue,191}, fill=none, line width=0.5mm]
\tikzstyle{DashedPurpleLine}=[-, densely dashed, draw={rgb,255: red,191; green,0; blue,191}, fill=none, line width=0.5mm]
\tikzstyle{DottedPurple}=[-, dotted, draw={rgb,255: red,191; green,0; blue,191}, fill=none, dotted, line width=0.5mm]
\tikzstyle{GreenLine}=[-, draw={rgb,255: red,80; green,200; blue,120}, fill=none, line width=0.5mm]
\tikzstyle{DashedGreenLine}=[-, densely dashed, draw={rgb,255: red,80; green,200; blue,120}, fill=none, line width=0.5mm]
\tikzstyle{DottedGreen}=[-, dotted, draw={rgb,255: red,80; green,200; blue,120}, fill=none, dotted, line width=0.5mm]
\tikzstyle{BrownLine}=[-, draw={rgb,255: red,210; green,105; blue,30}, fill=none, line width=0.5mm]
\tikzstyle{DashedBrownLine}=[-, densely dashed, draw={rgb,255: red,210; green,105; blue,30}, fill=none, line width=0.5mm]
\tikzstyle{DottedBrown}=[-, dotted, draw={rgb,255: red,210; green,105; blue,30}, fill=none, dotted, line width=0.5mm]
\tikzstyle{ArrowLineRight}=[-,  -{Stealth[scale=1.75]}, line width=0.25mm, scale=5]
\tikzstyle{ArrowLineRed}=[-, draw={rgb,255: red,191; green,0; blue,0},  -{Stealth[scale=1.75]}, line width=0.25mm, scale=5]
\tikzstyle{ArrowLineBlue}=[-, draw=blue,  -{Stealth[scale=1.75]}, line width=0.25mm, scale=5]
\tikzstyle{ArrowLinePurple}=[-, draw={rgb,255: red,191; green,0; blue,191},  -{Stealth[scale=1.75]}, line width=0.25mm, scale=5]
\tikzstyle{ArrowLineGreen}=[-, draw={rgb,255: red,80; green,200; blue,120},  -{Stealth[scale=1.75]}, line width=0.5mm, scale=5]
\tikzstyle{ArrowLineBrown}=[-, draw={rgb,255: red,210; green,105; blue,30},  -{Stealth[scale=1.75]}, line width=0.25mm, scale=5]

\tikzset{snake it/.style={decorate, decoration=snake}}
\tikzset{
dashstar/.style={
 dash pattern=on 5pt off 5pt,
 postaction={
  decorate,
  decoration={
   markings,
   mark=between positions 9pt and 1 step 10pt with {
     \node[color=red] {*};
   }
  }
 }
},
dashstarstar/.style={ 
 dash pattern=on 5pt off 10pt,
 postaction={
   decorate,
   decoration={
     markings,
     mark=between positions 10pt and 1
          step 15pt
           with {
            \node at (-2pt,0pt) {\pgfuseplotmark{star}};
            \node at (2pt,0pt) {\pgfuseplotmark{star}};
           }
   }
 }
}
}
\usepackage{pgfplots}
\pgfplotsset{compat=1.16}

\newcommand{\be}{\begin{equation}}
\newcommand{\ee}{\end{equation}}
\newcommand{\ba}{\begin{aligned}}
\newcommand{\ea}{\end{aligned}}

\begin{document}

\begin{flushright}
    CERN-TH-2026-030
\end{flushright}

\date{March 2026}

\title{A Matrix Theory Construction \\[3mm] of the IIA/IIB Wall}

\institution{CERN}{\centerline{$^{1}$CERN, Theoretical Physics Department, 1211 Meyrin, Switzerland}}

\authors{Ethan Torres\worksat{\CERN}\footnote{e-mail: \texttt{ethan.martin.torres@cern.ch}}
}

\abstract{In this note, we give a non-perturbative construction of a lightlike domain wall separating IIA and IIB string theories in 10D in the framework of discrete light-cone quantization (DLCQ). In this setting, generalizations of the BFSS conjecture relate the 10D flat space limit to matrix string theories (MSTs) for IIA and IIB. The former is equivalent to the large-$N$ limit of 2D Super Yang-Mills theory, while the latter is the large-$N$ limit of 3D ABJM theory with $\pm 1$ Chern-Simons levels. Our construction requires the string coupling to vanish at the location of the wall, and we show that BPS IIA $D0$-branes become non-BPS IIB $D0$-branes as they cross it, as anticipated in \cite{Heckman:2025wqd}.}


\maketitle




\section{Introduction}
What is the spectrum of branes in string theory? The extent to which we can answer this question is clearly a benchmark for our non-perturbative understanding of the theory. Moreover, it is a necessity if we wish to quantitatively understand vacua with little to no supersymmetry.

For 10D superstring theories in particular, much progress has been made on this question. As is well-known, BPS branes in these theories were discovered and classified in the mid-1990s and coincided with our understanding of strong coupling phase structure of 10D string theories. Additionally, non-BPS D-branes were discovered around the same time period, some of which are stable and carry K-theory charge.\footnote{Additionally, Type II non-BPS D-branes, which carry no $K$-theory charge in the traditional sense of \cite{Witten:1998cd}, can carry a $\mathbb{Z}_2$-valued charge as argued in \cite{Heckman:2025wqd}. For a review of non-BPS D-branes, see \cite{Sen:1999mg}.} In the subsequent decades, such insights were central in constructing string vacua (e.g. by the inclusion of branes wrapping cycles of the internal geometry), and for understanding quantum corrections to the low-energy effective field theory. However, the classification of branes in these settings is not fully understood.

More recently, new non-BPS branes have been conjectured to exist in 10D superstring theories. These include the R7-branes of Type II string theories \cite{Dierigl:2022reg,Dierigl:2023jdp}, a myriad of branes in heterotic string theories \cite{Kaidi:2023tqo,Kaidi:2024cbx}, and a domain wall between IIA and IIB \cite{Heckman:2025wqd}. These works have drawn inspiration from the Swampland Cobordism Conjecture \cite{McNamara:2019rup}. Namely, the existence of these branes are postulated in order to satisfy the Swampland constraint that certain quantum gravity cobordism classes must vanish. The existence of a IIA and IIB domain wall is then the simplest non-trivial prediction of \cite{McNamara:2019rup} in the sense that such a wall would separate pairs of theories which both have maximal supersymmetry and dimensionality (while still having different local physics).\footnote{We also note that a IIA/IIB wall domain wall was also conjectured to exist in \cite{Distler:2009ri}.}

A closely related motivation is that the existence of these branes is required from the expectation that quantum gravity theories satisfies charge completeness \cite{Banks:2010zn, Rudelius:2020orz, Heidenreich:2021xpr}.\footnote{Charge completeness is the statement that if a quantum gravity theory has a $p$-form gauge theory, then there must exist objects (or collections of objects) charged under all representations of the gauge theory. For a recent refinement of this idea, see \cite{Nevoa:2025xiq}. } For example, in Type IIA string theory left-moving spacetime fermion parity, $(-1)^{F_L}$, is a worldsheet global symmetry which lifts to a gauge symmetry in the 10D spacetime. A $\mathbb{Z}_2$ 0-form gauge theory in 10D must contain vortex objects with the associated monodromies as otherwise there would be a $\mathbb{Z}_2$ $8$-form global symmetry.\footnote{The topological operators for this $8$-form global symmetry are the Wilson lines for the $\mathbb{Z}_2$ connection.}

While there exist various pieces of evidence that these new non-BPS branes exist, non-perturbative constructions for them have so far remained elusive. Part of the difficulty is that, in contrast to non-BPS D-branes discovered long ago, these branes are not known to admit worldsheet CFT constructions which do not contain regions of strong string coupling.

In this note, we work in the setting of BFSS-like constructions of IIA and IIB string theories which conjecturally give non-perturbative definitions of these theories in 10D Minkowski space. After a brief review of the necessary facts on the discrete lightcone quantization (DLCQ) of Type II theories in Section \ref{sec:review}, we construct a lightlike domain wall\footnote{Constructions of lightlike domain walls connecting different string theories have been given previously in \cite{Hellerman:2006ff, Hellerman:2007fc}. However, the key physical difference with the IIA/IIB wall is that these constructions connect unstable string theory backgrounds to stable/less unstable ones via tracking the condensation of closed string tachyons. } in Section \ref{sec:main}. Section \ref{sec:discussion} includes our discussion and conclusions. Appendix \ref{app:TheOrb} includes more details on $\mathbb{Z}_2$ gauging of Green-Schwarz worldsheet CFTs.

\paragraph{\textit{Note added:}} While completing this work, the author was informed of \cite{Anastasi:2026cus} which constructs lightlike IIA/IIB walls in a different manner. We thank the authors for coordinating the arXiv submission.

\section{Brief Review of DLCQ of Type II String Theories}\label{sec:review}
The BFSS model \cite{Banks:1996vh} is a non-perturbative definition of M-theory in 11D Minkowski space from a 1D $U(N)$ SYM theory in the strict $N\rightarrow \infty$ limit.\footnote{For a recent presentation on the precise formulation of the correspondence with 1D gauge observables to 11D M-theory S-matrix elements see \cite{Maldcite}. } There is no shortage of quality reviews on this subject, but the author found  \cite{Dijkgraaf:1997vv,Banks:1999az,Taylor:2001vb} especially helpful, as well as the more recent references \cite{Maldcite,Lin:2025iir,Cho:2026jzs}. The general idea of BFSS is that the compactification of M-theory on a null circle (which can roughly be seen as an infinitely boosted spacelike circle) closely matches the physics of $N$ bound states of $D0$ branes in IIA. In M-theory lightcone coordinates
\begin{equation}\label{eq:lccoords}
    ds^2=-2dx^+dx^-+\sum^{9}_{i=1}(dx^i)^2, \quad \quad \quad x^\pm=\frac{1}{\sqrt{2}}(x^0\pm x^{10})
\end{equation}
one identifies $x^-\sim x^-+2\pi R_{11}$. The relevant on-shell momentum variables are related as
\begin{equation}
    p^+=-p_-=\frac{N}{R_{11}}, \quad \quad p^-=-p_+=\frac{(p_i)^2}{2p^+}
\end{equation}
and if we interpret $x^+$ as a time direction, then $p^-$ can be interpreted as the DLCQ Hamiltonian for a system of $N$ units of momentum along of the circle. From a IIA perspective, this system is a bound state of $N$ $D0$-branes which are the lightest degrees of freedom in the system in the $R_{11}\rightarrow \infty$ limit and are conjectured to capture all of the known properties of 11D M-theory physics. Much of this picture was further clarified from the holographic perspective of the $D0$-brane system \cite{Itzhaki:1998dd,Polchinski:1999br}.

Not long after the BFSS model was introduced, it was realized that DLCQ constructions also follow for M-theory on $T^n$ at least when $n\leq 5$. In the limit of $\mathrm{Vol}(T^n)\ll 1$ in 11D Planck units, this recovers IIA for $n=1$ and IIB for\footnote{Recall that M-theory on a $T^2$ with vanishing area is equivalent to IIB where the $M2$ brane winding is identified with momentum along one of the spatial directions of 10D Minkowski space \cite{Schwarz:1995dk, Aspinwall:1995fw, Schwarz:1995jq}, a foundational fact in F-theory \cite{Vafa:1996xn}.} $n=2$. The analogs of the Super Quantum Mechanics of the $N$ $D0$ theory can be obtained by acting with T-duality. As we review below, the IIA and IIB 10D flat spacetime limits require an infinite gauge coupling limit of the 2D and 3D SYM theories respectively. The former recovers the usual Matrix String Theory \cite{Motl:1997th, Dijkgraaf:1997vv}, while the latter can be shown to be equivalent to the CFT of $N\rightarrow \infty$ $M2$-branes, which is the ABJM theory at $k=1$. In what follows, we will simply call the former the IIA MST and the latter the IIB MST.\footnote{To the author's knowledge, the first proposal for the IIB MST appeared in \cite{Banks:1996my,Sethi:1997sw}, well before ABJM. In fact one of the most recent references on this subject is \cite{Gomis:2008cv} which appeared, incidentally, after the BLG-model \cite{Bagger:2007jr,Gustavsson:2007vu} on multiple $M2$-branes but still before ABJM. This means their method of treating a large-$N$ number of coincident $M2$-branes using BLG should, in retrospect, be replaced by $k=1$ ABJM CFTs at large-$N$ in light of \cite{Gauntlett:2008uf, nagy2008prolongationsliealgebrasapplications, Papadopoulos:2008sk}. }





\subsection{IIA from 2D $\mathcal{N}=(8,8)$ $U(N)$ SYM}
Reviewing the construction of \cite{Dijkgraaf:1997vv}, the DLCQ of IIA string theory is conjectured to be given by 2D $\mathcal{N}=(8,8)$ $U(N)$ SYM theory with worldvolume $\mathbb{R}\times S^1_R$. Let $\tau$ and $\sigma\sim \sigma +2\pi R$ be the coordinates of this cylinder, then the action is given by
\begin{equation}\label{eq:DVVaction}
    S=\int d\tau d\sigma \; \mathrm{tr}\left(-\frac{g^2_s}{4}(F_{\mu\nu})^2-\frac{1}{2}(D_\mu X^i)^2-\frac{1}{2}\theta^T\slashed{D}\theta+\frac{1}{g^2_s}[X^i,X^j]^2-\frac{1}{g_s}\theta^T \gamma_i[X^i,\theta] \right)
\end{equation}
where $X^i_{ab}$, $i=1,...,8$, and $\theta_{ab}=(\theta^{\Dot{\alpha}}_+, \theta^{\alpha}_-)^T_{ab}$ are $N\times N$ matrices which are in the adjoint representation of $U(N)$. The indices $\mu, \nu=0,1$ indicate the $\tau$ and $\sigma$ directions respectively. $X^i$, $\theta^{\Dot{\alpha}}_+$, and $\theta^{\alpha}_-$ are respectively in the $\mathbf{8}_v$, $\mathbf{8}_c$, and $\mathbf{8}_s$ representations of $Spin(8)_R$ global symmetry of the SYM theory, and $\pm$ denotes the 2D chirality of the Majorana-Weyl fermions. This note uses the opposite of the textbook conventions of \cite{Becker:2006dvp}, where $(+)$ here denotes left-moving chirality, while $(-)$ denotes right-moving chirality. The DLCQ variables are related to the gauge theory data as
\begin{equation}\label{eq:IIAvariables}
    g_s=\frac{1}{g_{YM}\ell_s}, \quad \quad P^+=\frac{N}{R}, \quad \quad  X^-\sim X^-+2\pi R
\end{equation}
where $X^\pm=\frac{1}{\sqrt{2}}(X^0\pm X^{9})$. While one can obtain a more conventional presentation of the SYM action by rescaling $(X^i,\theta)\rightarrow (g_s \ell_s X^i,g_s \ell_s\theta)$, the conventions of \eqref{eq:DVVaction} will make more clear that in the $g_s\rightarrow 0$ limit we obtain the Fock space of free IIA string theory. Indeed, as $g_s\rightarrow 0$, the gauge theory becomes infinitely strongly coupled and any commutators between $X$ and $\theta$ are energetically forced to vanish. Let $X^i_a:= X^i_{aa}$, $\theta_{a}:= \theta_{aa}$, then the action schematically reduces to
\begin{equation}
 -\sum^N_{a=1}\int d\tau d\sigma((\partial X^i_{a})^2+\theta^{\Dot{\alpha}}_{a, +}\slashed{\partial}_-\theta^{\Dot{\alpha}}_{a, +}+\theta^\alpha_{a, -}\slashed{\partial}_+\theta^\alpha_{a, -})
\end{equation}
which we recognize as simply $N$ copies of the IIA Green-Schwarz (GS) string action. After imposing the residual $U(N)$ gauge constraints, this is the $\mathcal{N}=(8,8)$ sigma model SCFT with orbifold target space $\mathrm{Sym}^N(\mathbb{R}^8)=(\mathbb{R}^8)^{\otimes N}/S_N$. Twisted sectors are labeled by conjugacy classes of the symmetric group, $S_N$, which are the charges which respect to the $\mathrm{Rep}(S_N)$ 0-form quantum symmetry. IIA strings with longitudinal momentum $P^+=n/R$ (for $n\leq N$) are associated with the conjugacy class $[g]$ where $g=(i_1 \; i_2 \; ...\; i_n)$ with distinct integers $1\leq i_j\leq N$ is a cyclic permutation of $n$ elements.

In the large-$N$ limit, the so-called long strings, which have fixed $P^+$ as $R\rightarrow \infty$, are the IIA fundamental strings. It was also shown in \cite{Dijkgraaf:1997vv}, that the leading $g_s$ perturbation is given by an irrelevant deformation of the SCFT which precisely matches the interaction vertex in lightcone string perturbation theory.


\subsection{IIB from 3D $\mathcal{N}=8$ ABJM}
The MST for IIB is conjectured to be given by a torus compactification of 3D $\mathcal{N}=8$ ABJM CFT \cite{Aharony:2008ug} with gauge group $U(N)_{1}\times U(N)_{-1}$ where the subscripts denote the Chern-Simons levels. The worldvolume of the theory is $\mathbb{R}\times T^2_{R_1,R_2}$ with spatial coordinates $\sigma_{i}\sim \sigma_{i}+2\pi R_i$ and correspond the the IIB DLCQ variables as
\begin{equation}\label{eq:IIBvariables}
   g_s=\frac{R_2}{R_1}, \quad \quad P^+=\frac{N\ell_s}{R_1R_2}, \quad \quad  X^-\sim X^-+2\pi \frac{R_1 R_2}{\ell_s}
\end{equation}
Here we have taken $T^2_{R_1,R_2}$ to be a rectangular torus (i.e. $C_0=0$) for simplicity. From the point of view of the coincident $M2$-brane description of the MST, only the $3D$ $\mathcal{N}=8$ CFT degrees of freedom are relevant in the $R_1R_2\rightarrow \infty$ limit which is the conjectured dual description of IIB in 10D Minkowski space. Superconformal symmetry implies that there is a $Spin(8)_R$ R-symmetry with the usual Poincar\'e supercharges in the $(\mathbf{2},\mathbf{8}_s)$ representation of $Spin(1,2)\times Spin(8)_R$.\footnote{For a detailed study of this enhancement of $k=1$ ABJM theory see \cite{Gustavsson:2009pm}.} The free IIB limit is $R_2\rightarrow 0$ while maintaining $R_1R_2\rightarrow \infty$. To get a handle on it, let us first consider finite $R_1R_2$. At a generic point in the moduli space we have a 3D $U(1)^N$ $\mathcal{N}=8$ SYM theory which has manifest $Spin(7)$ symmetry. If we dualize the vectors to periodic scalars, $dX_a^8:=*dA_a$, then in the $R_1R_2\rightarrow \infty$ limit we have $g_{U(1)_a}\rightarrow \infty$ which corresponds to a decompactification of the $X^8$ target direction. The non-trivial Chern-Simons level ABJM theory is required for the moduli space to be $(\mathbb{R}^8)^N/S_N$ in this limit. This leads to an action
\begin{equation}
    -\sum^N_{a=1}\int d\tau d\sigma_1 d\sigma_2 \left((\partial_\mu X_a^i)^2+\Theta_a \slashed{\partial} \Theta_a\right)
\end{equation}
If we also take $R_2\rightarrow 0$, then the $3D\rightarrow 2D$ reduction is $N$ copies of the IIB lightcone GS action
\begin{equation}
    -\sum^N_{a=1}\int d\tau d\sigma_1((\partial X^i_{a})^2+\theta^\alpha_{a, +}\slashed{\partial}_-\theta^{\alpha}_{a, +}+\theta^\alpha_{a, -}\slashed{\partial}_+\theta^\alpha_{a, -})
\end{equation}
As we review in Appendix \ref{app:TheOrb}, the IIB GS worldsheet is equivalent to the IIA GS worldsheet by gauging the left-moving fermion parity of the latter,
\begin{equation}\label{eq:FLstudios}
    (-1)^{F_L}_{IIA}: \theta^{\Dot{\alpha}}_{+}\rightarrow -\theta^{\Dot{\alpha}}_{+}.
\end{equation}
This means that the IIB MST at $g_s=0$ can be seen as an orbifold of the $\mathrm{Sym}^N(\mathbb{R}^8)$ theory by simultaneously gauging the $N$ $\mathbb{Z}_2$-symmetries
\begin{equation}\label{eq:FLstudiosmia}
 (-1)^{F_L}_{IIA,a}:   \theta^{\Dot{\alpha}}_{a,+}\rightarrow -\theta^{\Dot{\alpha}}_{a,+}
\end{equation}
Equivalently, this can be seen as starting with $N$ copies of the IIA GS lightcone worldsheet CFT, gauging by all of the $\mathbb{Z}_2$ symmetries \eqref{eq:FLstudiosmia}, then gauging by $S_N$. We can schematically summarize these 2D CFT relations as
\begin{align}
    \mathrm{MST}_{A}(g_s=0)&=\frac{(\mathrm{GS}_A)^{\otimes N}}{S_N} \\
    \mathrm{MST}_{B}(g_s=0)&=\frac{(\mathrm{GS}_A/\mathbb{Z}_2)^{\otimes N}}{S_N}=\frac{(\mathrm{GS}_B)^{\otimes N}}{S_N}
\end{align}
where $GS_{A/B}$ denotes a single copy of the IIA/IIB GS lightcone worldsheet CFT, the $\mathbb{Z}_2$ gauging is left-moving fermion parity, and we made explicit the $S_N$-gauging in the symmetric orbifold.

Finally, we note that the reader may be more familiar with the IKKT proposal for a non-perturbative definition of Type IIB string theory from a $0+0D$ matrix integral \cite{Ishibashi:1996xs}. There is no contradiction with the IIB MST in the same way that the 11D flat space limit of M-theory can equally be defined by taking the flatspace limit of $AdS_4\times S^7$ or $AdS_{7}\times S^4$ (e.g. recent consequences of this correspondence were explored in \cite{Chen:2026fpe, Hristov:2026tde}). In the following section, our proposal for the IIA/IIB wall will require a Hamiltonian description on both sides which precludes the use of a $0+0D$ theory for this purpose.


\section{IIA/IIB Wall}\label{sec:main}
We now turn to the main subject of this note which is a non-perturbative construction of a domain wall separating IIA and IIB string theories. We start with two observations of the Type II matrix theory constructions. First, notice that IIA or IIB MSTs can each be generalized from taking constant asymptotic boundary values of the dilaton, $\phi=\log (g_s)$, to any smooth function $\phi(X^+)$. From \eqref{eq:IIAvariables} and \eqref{eq:IIBvariables}, we see that this amounts to a varying $R(\tau)$ for the IIA MST and $\frac{R_2}{R_1}(\tau)$ for the IIB MST respectively. Although we are interested in the strict $N\rightarrow \infty$ limit, we remark that for finite $N$, this has a holographic interpretation in the respective $D1$ and $M2$ brane systems as setting the value for the string dilaton at the AdS-like boundary. This becomes $\phi(X^+)|_{X^-=\pm \infty}$ in the flatspace limit.\footnote{For a review of finite $N$ holographic solutions of these systems, see \cite{Lin:2025iir}.} The second observation, as shown at the end of Section \ref{sec:review} and in more detail in Appendix \ref{app:TheOrb}, is that when the asymptotic coupling vanishes, $g_s=0$, the IIB MST is a $\mathbb{Z}_2$-orbifold of the IIA MST.

\begin{figure}[t]
    \centering
    \includegraphics[scale=0.45]{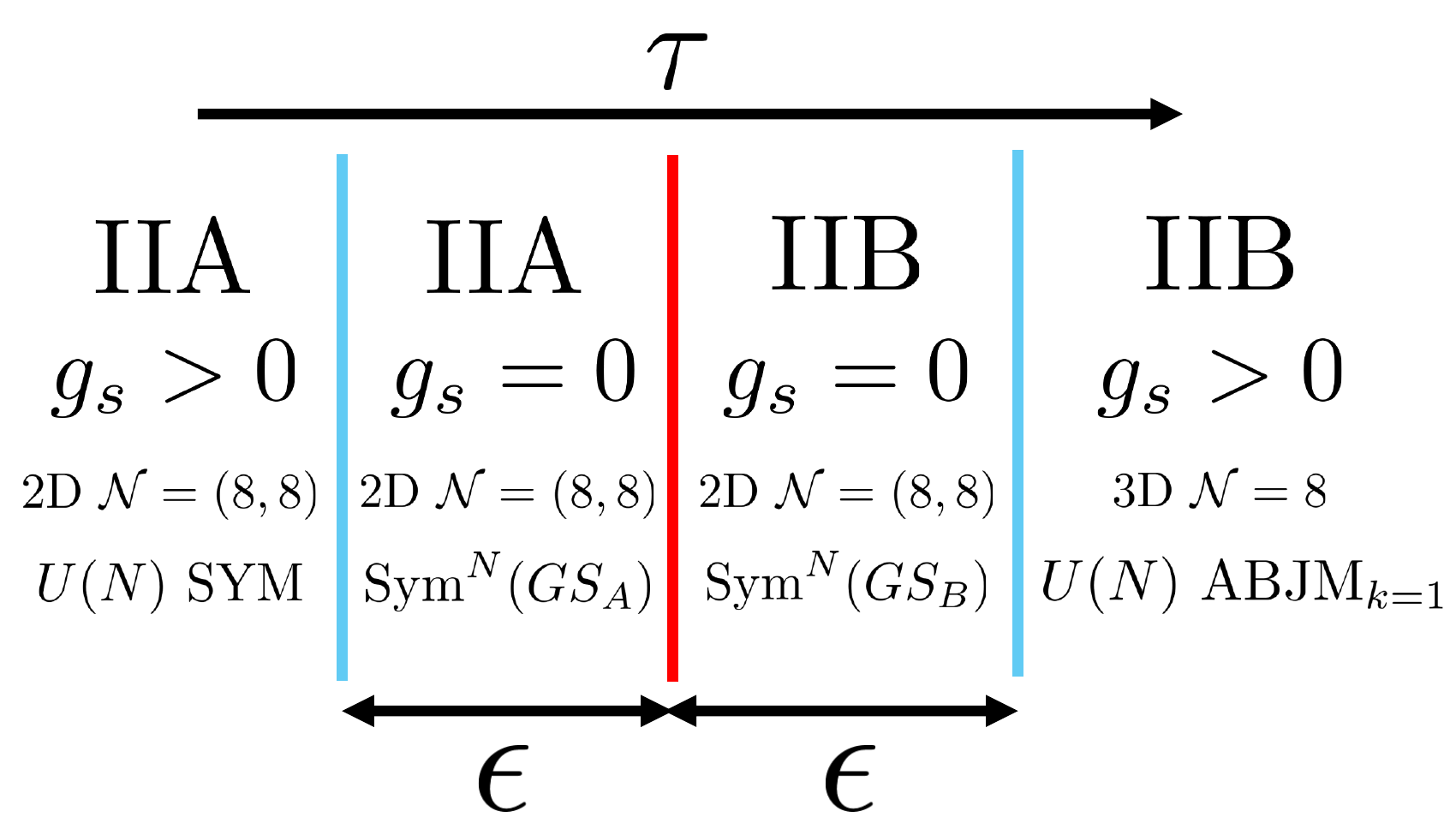}
    \caption{Matrix string theory construction of IIA/IIB wall. The $(-1)^{F_L}_{IIA,a}$ $\mathbb{Z}_2$-symmetries of the symmetric orbifold CFT in the IIA $g_s=0$ region are gauged to the right of the red line. The $g_s>0$ regions correspond to turning on certain irrelevant deformations of the 2D symmetric orbifold SCFTs. On the IIA side, this irrelevant deformation corresponds to flowing away from $g_{YM}^2=\infty$ to finite gauge coupling. On the IIB side, the irrelevant deformation corresponds to a 2D $\rightarrow$ 3D decompactification.}
    \label{fig:MSTAB}
\end{figure}

\begin{figure}
    \centering
    \includegraphics[width=10cm]{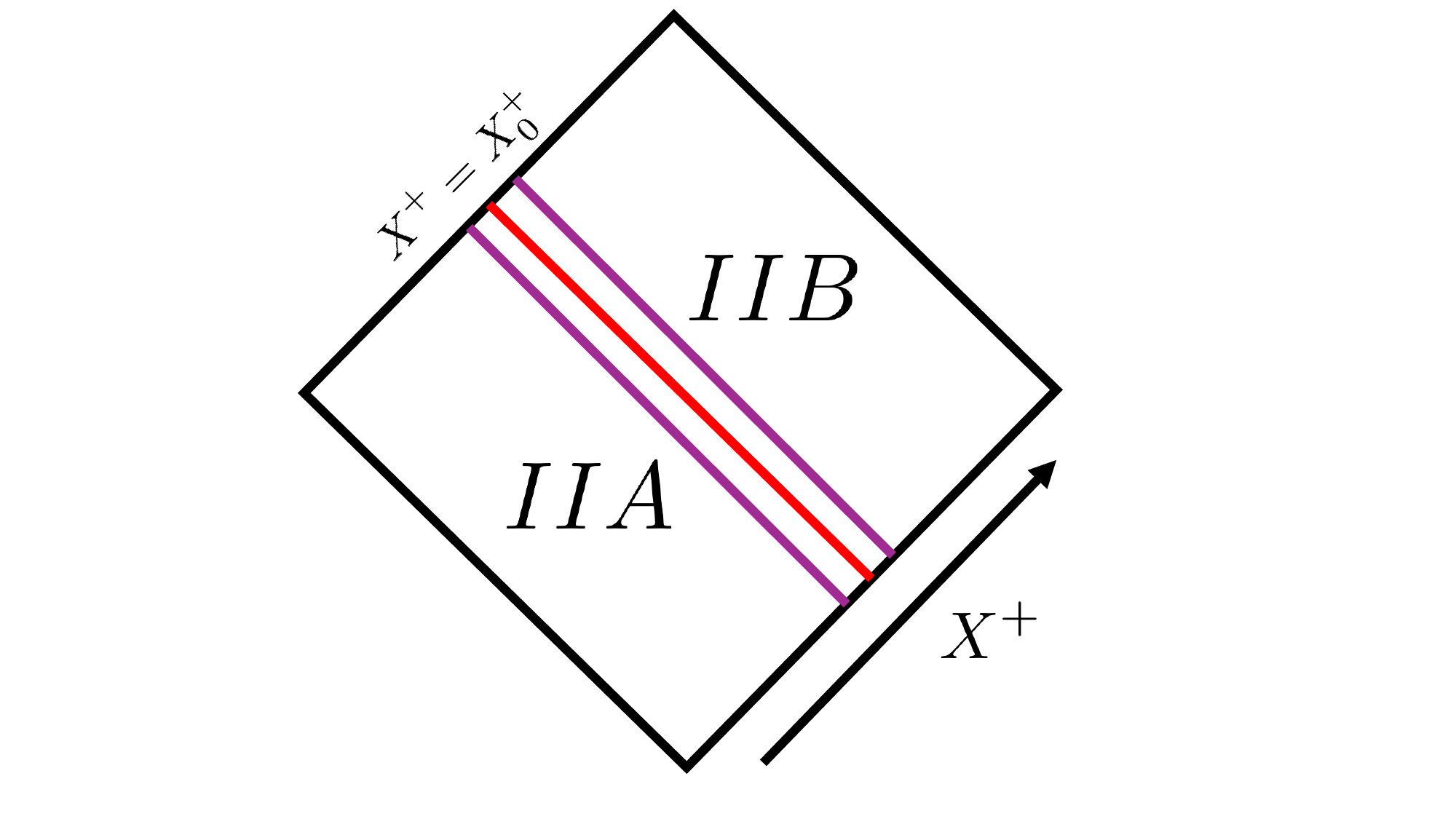}
    \caption{10D Minkowski spacetime illustration of IIA/IIB wall located at $X^+=X^+_0$. In DLCQ coordinates, this is located at $\tau=\tau_0$. Between the purple lines indicates a region where the Einstein frame curvature is of the order of the string mass scale. See section \ref{ssec:spacetimepic} for more details. }
    \label{fig:PenroseAB}
\end{figure}

\textbf{IIA/IIB Wall Definition:} Let $g_{s}(\tau)=0$ for all $\tau$ the some neighborhood $(\tau_0-\epsilon, \tau_0+\epsilon)$, then one can form a lightlike IIA/IIB domain wall at $\tau_0$ by gauging the $\mathbb{Z}^{\times N}_2$  symmetry generated by the $(-1)^{F_L}_{IIA, a}$ elements \eqref{eq:FLstudiosmia} in the region $\tau\in [0,\tau_0+\epsilon)$.\footnote{Note that this is a symmetric orbifold analog of the half-space gauging considered on fundamental GS string worldsheets piercing the IIA/IIB wall considered in \cite{Heckman:2025wqd}. For early references on half-space gauging of discrete symmetries see \cite{Kaidi:2021gbs, Choi:2021kmx,Roumpedakis:2022aik}.} See Figure \ref{fig:MSTAB} for an illustration of the MST setup. This is well-defined in the limit $\epsilon\rightarrow 0$, where $g_s$ vanishes at an isolated value $\tau=\tau_0$, because along the free MST surface $\tau=\tau_0$, the half-interval gauging of $(-1)^{F_L}_{IIA,a}$ in $(\tau_0-\epsilon, \tau_0+\epsilon)$ reduces to a codimension-1 gauging at the $\tau=\tau_0$ line. Gauging a symmetry only along a codimension-1 submanifold is known as a codimension-$k$ condensation defect \cite{Roumpedakis:2022aik} which is a well-defined object for anomaly-free discrete symmetries. See Figure \ref{fig:PenroseAB} for a spacetime illustration of this IIA/IIB wall located at the hypersurface $X^+=X^+_0$ ($\tau=\tau_0$ in MST coordinates).



\subsection{Objects Crossing the Wall}\label{ssec:crossing}

With a non-perturbative definition of the IIA/IIB wall in hand, we now comment on the effect the wall has on various Type II objects. In the notation of Figure \ref{fig:MSTAB}, we will work in the middle two regions at finite $\epsilon>0$ for ease of exposition.

By construction, a fundamental IIA string becomes a fundamental IIB string as it crosses the wall. However, one can still ask: what is the fate of the individual string states with $(-1)^{F_L}=-1$ as they cross the wall\footnote{We will occasionally drop the IIA and IIB subscripts on $(-1)^{F_L}$ when the interpretation is clear.}? From Appendix \ref{app:TheOrb}, we know from the orbifolding procedure that the left-moving Ramond sector of IIA is projected out of the Hilbert space for $\tau>\tau_0$, while the left-moving Ramond states of IIB will arise in the twisted sector Hilbert space. We also saw from Appendix \ref{app:TheOrb}, that these $(-1)^{F_L}_{IIA}=-1$ states of IIA still exist as states albeit in a \textit{defect} Hilbert space, see Figure \ref{fig:stateop}. The defect in question is the topological line defect which generates the quantum symmetry of the $(-1)^{F_L}_{IIA}$ gauging, which is $(-1)^{F_L}_{IIB}$. In the formalism of GS lightcone perturbation theory \cite{Green:1987mn}, these defect Hilbert spaces are not included so these do not lead to propagating states in this setting.

From the $g_s=0$ MST perspective however, there is a priori no reason to discard such states as the defect Hilbert space data is part of the CFT data. Indeed we can argue explicitly that they gain a non-perturbative correction to their masses when $g_s>0$. To see this, it will be more straightforward to flip the setup $\tau\rightarrow -\tau$ and instead consider defect Hilbert space states for Type IIA MST. In other words, these are $(-1)^{F_L}_{IIB}=-1$ states in the usual (i.e. non-defect) Hilbert space on the IIB side.


To define the IIA defect Hilbert space, we must first generalize the $(-1)^{F_L}_{IIA}$ symmetry action at $g_s=0$ \eqref{eq:FLstudios} to $g_s>0$ since the former was only given in terms of diagonal matrix components of the IIA MST fields. Of the $N$ $\mathbb{Z}_2$ generators \eqref{eq:FLstudiosmia}, only the diagonal $\mathbb{Z}_2$ action is preserved at finite $g_s$. This lift takes the form
\begin{equation}
    (-1)^{F_L}_{IIA}: \; \; X^i\rightarrow +(X^i)^\mathrm{T}, \quad \quad \theta^{\Dot{\alpha}}_+\rightarrow -(\theta^{\Dot{\alpha}}_+)^\mathrm{T}, \quad \quad \theta^{\alpha}_-\rightarrow +(\theta^{\alpha}_-)^\mathrm{T}, \quad \quad A_\mu\rightarrow -(A_\mu)^\mathrm{T}
\end{equation}
which means that IIA MST states in the defect Hilbert space have the monodromy
\begin{align}
  &  X^i(\sigma+2\pi R)=(X^i(\sigma))^\mathrm{T}, \quad \quad \theta^{\Dot{\alpha}}_+(\sigma+2\pi R)=-(\theta^{\Dot{\alpha}}_+(\sigma))^\mathrm{T}, \\
&  \theta^{\alpha}_-(\sigma+2\pi R)=(\theta^{\alpha}_-(\sigma))^\mathrm{T}, \quad \quad A_\mu(\sigma+2\pi R)=-(A_\mu(\sigma))^\mathrm{T}\label{eq:Aaction}
\end{align}
We know from the IIB side that the ground state energy for this sector vanishes at $g_s=0$ simply because the $\mathbb{Z}_2$ gauging commutes with the Hamiltonian. The main point then is that the ground state energy of this sector is no longer protected by supersymmetry when $g_s>0$. This is essentially because this sector is a chiral Scherk-Schwarz compactification (composed with charge conjugation) and is shown explicitly to have a non-zero Casmir energy in Appendix \ref{app:TheOrb}. The reason why $g_s>0$ is crucial is that when $g_s=0$ the symmetric orbifold CFT has a level-matching condition which can be violated when $g_s>0$ as shown in \cite{Dijkgraaf:1997vv}.

In summary, this means that IIB states with $(-1)^{F_L}=-1$ gain a string scale mass when crossing to the IIA side. A similar argument can also be made in the opposite direction. A more refined estimate of such masses, as well as engineering scenarios where some of these are fine-tuned to vanish would be interesting to pursue in future work.



\begin{figure}
    \centering
    \includegraphics[width=14cm]{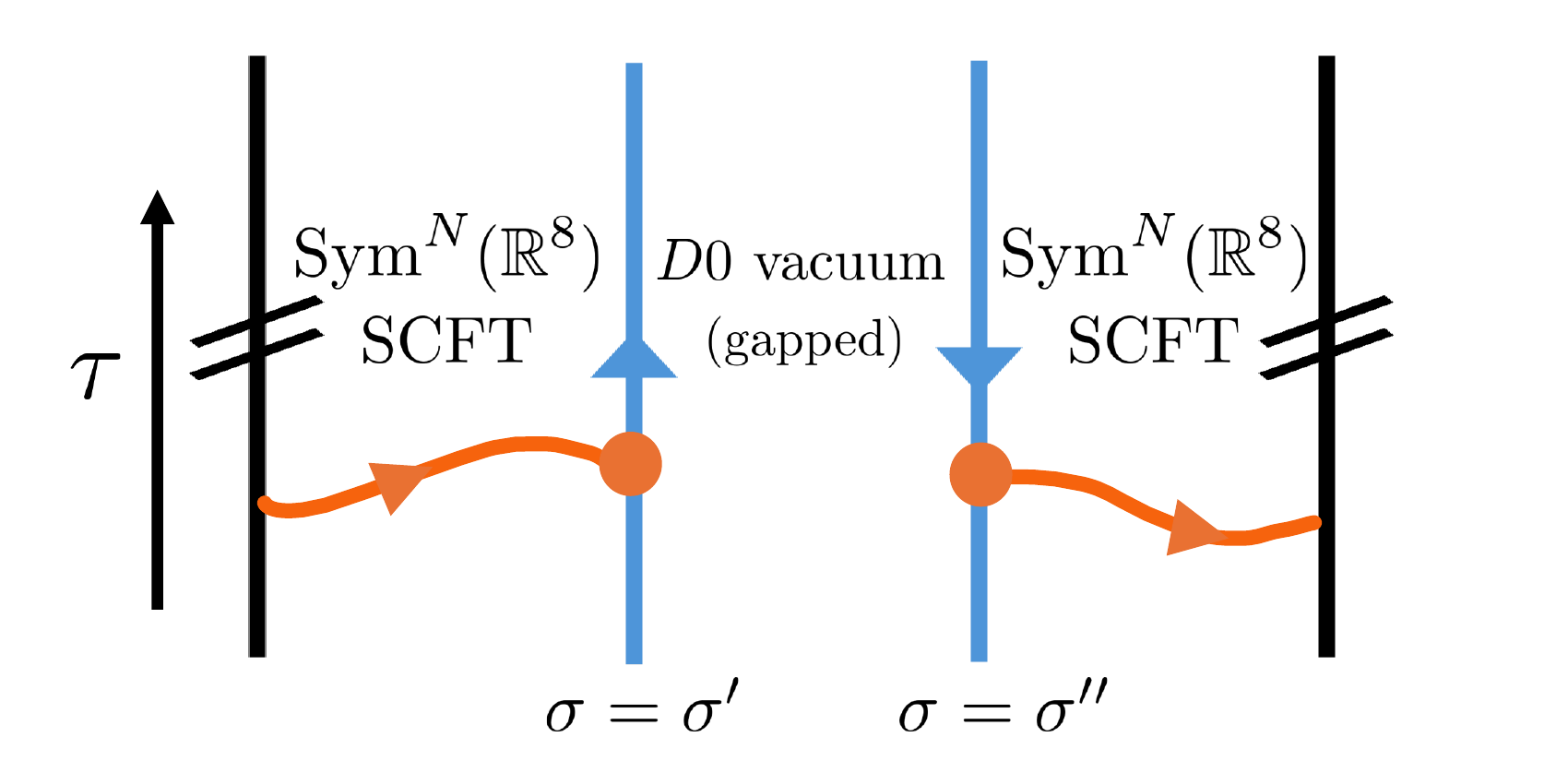}
    \caption{Symmetric orbifold configuration on $\mathbb{R}_\tau\times S^1_\sigma$ dual to a BPS IIA $D0$-brane localized along $X^-$. We have conformal interfaces (blue lines) separating the symmetric orbifold from the gapped $D0$-brane vacuum. The black lines are identified. The orange lines represent a single GS string with Dirichlet boundary conditions on the interfaces. These conformal interfaces are the IR limit of $U(N)$ Wilson lines in conjugate representations. Not pictured is are the overall gappless degrees of freedom in the $D0$ vacuum from the spin-0 and spin-1/2 components of the overall $U(1)$ vectormultiplet.}
    \label{fig:symbc}
\end{figure}

It was argued in \cite{Heckman:2025wqd} that BPS D-branes should become non-BPS D-branes when crossing the IIA/IIB wall. Here we study fate of IIA $D0$-branes crossing the IIA/IIB wall to understand if it indeed becomes a non-BPS $D0$-brane on the IIB side. In the language of MST, the $D0$-charge is $Q_{D0}=\int_{S^1_R} \mathrm{Tr}(F)$ \cite{Dijkgraaf:1997vv}. This follows from the standard duality argument of IIA MST: the 2D $U(N)$ SYM describes $N$ $D1$s in a dual IIB frame and the flux $\mathrm{Tr}(F)$ sources $D(-1)$-branes which map to $D0$s under T-duality. In terms of the decomposition, $U(N)=(SU(N)\times U(1))/\mathbb{Z}_N$, a $Q_{D0}=1$ background corresponds to turning on a $1 \; \mathrm{mod}\; N$ flux in the $SU(N)$ factor and a $-1/N$ flux in the $U(1)$ factor. As explained in \cite{Dijkgraaf:1997vv}, such a flux sector has a diverging ground state energy in the $g_s\rightarrow 0$ limit, and indeed there is strong evidence that this $SU(N)$ flux vacuum is gapped \cite{Kologlu:2016aev}. To extract the open string spectrum of the $D0$-brane, we consider a modification of the original construction of the $D0$-brane in \cite{Dijkgraaf:1997vv} to one localized along the $X^-$-direction. This can be done by sourcing the $Q_{D0}=1$ flux by a Wilson line at some $\sigma=\sigma'$ in the $\mathbf{N}_{-1/N}$ representation of $(SU(N)\times U(1))/\mathbb{Z}_N$, as well as a Wilson line in the conjugate representation, $\overline{\mathbf{N}}_{+1/N}$, located at some $\sigma=\sigma''$. See Figure \ref{fig:symbc} for an illustration. These Wilson lines are conformal interfaces in the $g_s\rightarrow 0$ limit and must be completely reflective in the sense of \cite{Bachas:2001vj}. Therefore the choice of boundary conditions for the symmetric orbifold fields are either Dirichlet or Neumann. While both seem to in principle be allowed, we choose Dirichlet boundary conditions since we are interested in engineering the $D0$-brane massless modes.\footnote{Other choices are most likely equivalent to considering bound states of various Type IIA BPS and non-BPS $D$-branes with the $D0$, but we leave this refined understanding of the IIA MST dictionary to future work.} The $S_N$-invariant boundary conditions take the form \cite{Lambert:1999id}
\begin{equation}
    \partial_+ X^i_{a}|_{\sigma=\sigma', \sigma''}=-\partial_- X^i_{a}|_{\sigma=\sigma', \sigma''}, \quad \quad \theta^{\Dot{\alpha}}_{a,+}|_{\sigma=\sigma',\sigma''}=\Gamma_0  \theta^{\alpha}_{a,-}|_{\sigma=\sigma',\sigma''}
\end{equation}
where $\Gamma^0$ is a target spacetime gamma matrix. Symmetric orbifold states on this spatial interval then describe the $0$-$0$ open string sector in the GS-formalism with various values of $p^+$.

Since the $(-1)^{F_L}$ gauging turns BPS $Dp$-boundary conditions into non-BPS $Dp$-boundary conditions \cite{Heckman:2025wqd, Sen:1999mg}, we see that bringing the configuration of Figure \ref{fig:symbc} across the IIA/IIB wall will simply be a non-BPS $D0$-brane in Type IIB string theory. We also see from the action on the $U(N)$ gauge potential \ref{eq:Aaction} that $Q_{D0}$ is not a conserved quantity on the IIB side.


\subsection{Spacetime Picture}\label{ssec:spacetimepic}
We now briefly describe the spacetime picture of the IIA/IIB wall as constructed in the MST setup. Recall that the function $g_s(\tau)$, given in terms of IIA and IIB MST variables in \eqref{eq:IIAvariables} and \eqref{eq:IIBvariables}, controls the asymptotic value of the dilaton as a function of $X^+$. Given that the only condition we have on the smooth function $g_s(\tau)$ is that $g_s(\tau_0)=0$,\footnote{Here we take $\epsilon=0$.} there are a large plurality of IIA/IIB walls depending on what choice one makes. The MST formalism in particular allows one to choose profiles that do not have to satisfy an equation of motion as the, necessarily, non-conformal deformation of the string worldsheet is supplied with a non-perturbative definition. Notice also that because the $g_s=0$ is at infinite distance in the dilaton moduli space, the semiclassical description of the physics must necessarily break down, which indeed happens to due to the infinite tower of string excitations becoming massless in the Einstein frame.

Additionally, in order for the wall to not lie on the asymptotic boundary of Minkowski space, the curvature of the spacetime must reach the string scale close to the wall and, moreover, the function $g_s(\tau)$ cannot simply be set to a solution to the dilaton equations of motion. To illustrate this, consider the metric ansatz
\begin{equation}
    ds^2=e^{f}(-dX^+dX^-)+e^{-f/4}\left(\sum^8_{i=1}(dX^i)^2\right)
\end{equation}
where we also take $f=f(X^+)$, $\phi=\phi(X^+)$, and $\phi_0:=\phi(\pm \infty)$. The equations of motion for the dilaton and metric are simply
\begin{align}
    &\partial_{+}^2\phi=0 \\
    & R_{00}=R_{99}=\partial_{+}^2f-2(\partial_{+} f)^2=\kappa (\partial_{+} \phi)^2
\end{align}
where $\kappa=e^{\phi_0}$ in $\ell_s=1$ units, and we have used the fact that the scalar curvature of the ansatz metric vanishes. We see that while a lightlike linear dilaton profile can solve the equations of motion, such a solution will never reach the necessary $g_s=0$ condition used to place the IIA/IIB wall. Also, since the IIA/IIB wall region requires $|\partial_+ \phi|\rightarrow \infty$ in its neighborhood, we indeed see that the Ricci curvature will leave the realm of effective field theory. This is the purple region in Figure \eqref{fig:PenroseAB}.

If we for instance consider the asymptotic string coupling profile
\begin{equation}\label{eq:cases}
   g_s(\tau)=\begin{cases}
			g_{s,0}\left(1-\exp(\frac{\tau^2}{\tau^2-c^2})\right) & |\tau|\leq c\\
            g_{s,0} & |\tau|\geq c
		 \end{cases}
\end{equation}
then the Ricci curvature very quickly grows to the order of string scale in the region $-c \leq \tau\leq +c$.\footnote{Note that this parameter $c$ is not to be confused with $\epsilon$ from earlier.} A IIA/IIB wall placed at $\tau\; (=X^+)=0$ is can be described by Type II supergravity with constant dilaton/metric outside of this region, and requires the (irrelevant deformation of) the symmetric orbifold CFT description in the interior. This situation is not too dissimilar to the holographic dual solutions of $Dp$-branes worldvolume theories for $p\leq 2$ which always contain regions of the bulk spacetime which cannot be described by a gravitational EFT \cite{Itzhaki:1998dd}.

\begin{figure}
    \centering
    \includegraphics[width=12cm]{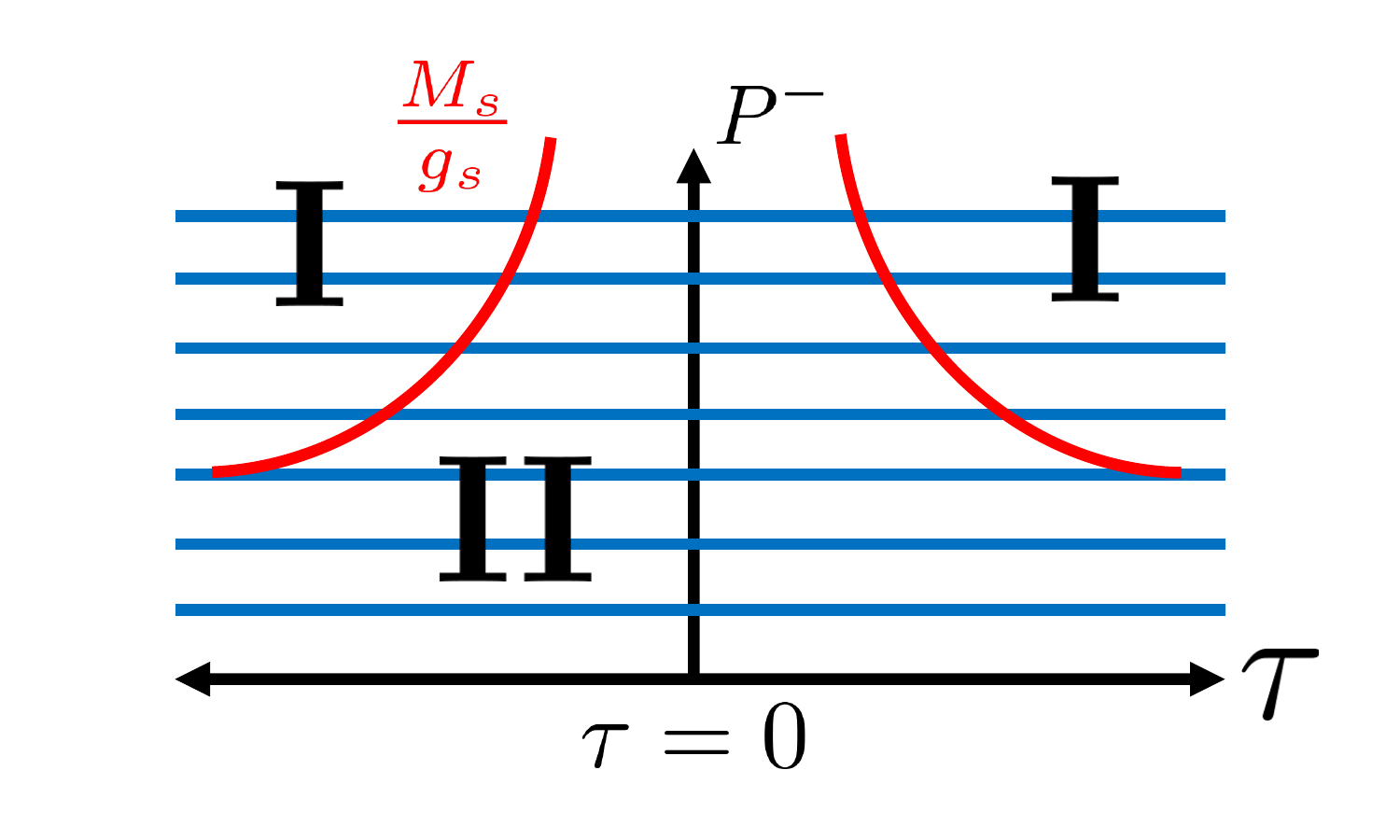}
    \caption{Depiction of validity of the 2D CFT treatment of IIA/IIB states near the domain wall at $\tau=X^+/\ell^2_s P^+=0$. Here the blue lines denote the masses of  IIA or IIB Fock space states for a fixed total $P^+$, which is a valid description below the red line $P^-<M_s/g_s(\tau)$. This means that the transmission of string states in region II across the wall can be treated in (time-dependent) irrelevant CFT perturbation theory, while string states in region I requires a non-perturbative treatment of the 2D SYM.}
    \label{fig:wave}
\end{figure}

\paragraph{Transmissibility and Finite Energy of Wall:} Notice that because the half-space gauging interface is topological, the IIA/IIB wall is tensionless in the $g_s=0$ core region. This means that the total stress-energy of the wall comes solely from the varying dilaton profile transverse to it. One cannot use the Type II supergravity EFT since it is not valid in the region where $|\nabla \phi|$ is large. However, we argue that the symmetric orbifold picture of the wall allows us to conclude that the wall is transmissible and of finite stress-energy. Recall that when $0<g_s\ll 1$, the IIA/IIB MSTs can be presented as irrelevant deformations of $\mathrm{Sym}^N(\mathrm{GS}_{A/B})$ where the irrelevant deformation to the Lagrangian takes the form\footnote{The DVV interaction was originally argued in the IIA MST, but their argument equally works for the IIB MST.} $\frac{g_s}{M_s}V_{\mathrm{int}}$ \cite{Dijkgraaf:1997vv} where $V_{\mathrm{int}}$ has dimension $(\frac{3}{2},\frac{3}{2})$. This means that the effect of this interaction on states with $P^-\ll M_s/g_s$ can be treated perturbatively.\footnote{Technically only first order perturbation theory is valid without including additional $g^n_s$ terms \cite{Dijkgraaf:2003nw}.} See figure \ref{fig:wave} for a depiction. In the context of the IIA/IIB domain wall, the effect of a $\tau$-dependent $g_s$ from the perspective of, say, graviton states can be treated using time-dependent perturbation theory. This clearly makes graviton radiation transmissible across the wall as long as $P^-\ll M_s/g^{max.}_s$. Moreover, since the effect of the irrelevant coupling can be adiabatically switched on, the stress-energy of the wall as measured from low-energy gravitons is adiabatically connected to 0, so we conclude that the stress-energy of our IIA/IIB wall must be finite. While we leave a precise calculation of this stress-energy as a functional of $g_s(\tau)$ for future work, we can conclude that our domain wall satisfies the Swampland Cobordism Conjecture prediction \cite{McNamara:2019rup} of a finite energy domain wall between IIA and IIB. Additionally, for each finite value of $N$, such a 2D/3D gauge theory interface precisely engineers a cobordism class-killing domain wall in an AdS-like holographic setting \cite{Ooguri:2020sua}.

\paragraph{If $g_s\ll 1$ Throughout Why Not Use String Perturbation Theory?}
In our construction of the IIA/IIB wall, any asymptotic value of $g_s$ when $|X^+|\rightarrow \infty$ was allowed, so one may wonder why, if this asymptotic value is taken to be small, a similar construction to ours has not appeared in string perturbation theory. The intuitive answer is that spatially varying $g_s$ in a manner which doesn't satisfy equations of motion is necessarily a non-conformal deformation of the perturbative worldsheet CFT. From the the 2D/3D MST perspective, this is simply sourced by a varying UV parameter which affects the strings in the IR limit, meaning that we have a well-defined non-perturbative completion of this non-conformal worldsheet. Indeed, the introduction of potential to the worldsheet CFT triggering a RG-flow was important in the IIA/IIB proposal of \cite{Anastasi:2026cus}, and it would be interesting to understand the perturbative/non-perturbative consistency.

\paragraph{IIA and IIB Indistinguishability at $g_s=0$}
While our construction requires $g_s=0$ at the location of the wall, this is by no means evidence that there does not exist a construction of the IIA/IIB wall where $g_s\neq 0$ at the wall. However, one can plausibly see why $g_s=0$ is conceptually pleasing since the Fock spaces of IIA and IIB string theories (which capture the whole theory when $g_s=0$) are finite gaugings of each other. When field theories satisfy this property (e.g. $SU(N)$ vs $PSU(N)$ pure gauge theories) they are locally indistinguishable, and it would be interesting to determine whether such an indistinguishability is required at the location of any domain wall between IIA or IIB, or domain walls between different string theories more generally.


\section{Discussion and Outlook}\label{sec:discussion}
In this note, we have given a non-perturbative definition of the IIA/IIB wall and initiated a study of its basic properties. It would be interesting to understand if similar constructions exist for $R7$-branes in Type II string theories. Indeed, it seems reasonable that a least light-like $R7$-branes can be constructed by defining the brane as a local operator in the MST theory which has a $(-1)^{F_L}$ topological symmetry line attached. Presumably, different choices of operators would correspond to different excited states of the $R7$. Matrix theory is a particularly appealing method to non-perturbatively define $R7$-branes since this is typically made difficult by the fact that well-studied AdS vacua often include fluxes which spontaneously break the gauge symmetry associated with the $R7$. For instance, the $F_5$-flux on Type IIB on $AdS_5\times S^5$ is not invariant under $(-1)^{F_L}$ nor $\Omega$, whereas even finite $N$ MSTs have purely geometric holographic duals \cite{Lin:2025iir}. It would be interesting to make contact with the $R7$-brane gravitational solutions found recently in \cite{Cavusoglu:2026xiv}.

Given that there exists BFSS-like constructions for heterotic string theory \cite{Danielsson:1996es, Motl:1996xx, Kachru:1996nd, Lowe:1997fc, Kim:1997uv, Banks:1997it, Banks:1997zs, Lowe:1997sx, Horava:1997ns}, it would also be interesting to pursue non-perturbative constructions of non-BPS branes of heterotic theories \cite{Kaidi:2023tqo,Kaidi:2024cbx}. Also, one may consider non-BPS branes in asymmetric orbifold backgrounds as these also admit matrix theory constructions \cite{Aharony:2007du}.

Another line of investigation is understanding how to define finite size IIA/IIB wall bubbles. Such objects would necessarily be unstable but would nevertheless be interesting objects that could arise in the non-perturbative S-matrix of Type II string theories. An observation we have made that may be useful in this pursuit is that the left-moving Ramond states of Type IIB remain as non-perturbative albeit massive states on the IIA side and vice-versa. Additionally, assuming the existence of such walls, it was conjectured by Sen recently in \cite{Sen:2025iuf}, that if such a bubble exists and does not lie behind an event horizon, then the tension must vanish at some $g_s$. If one takes this statement to mean $g_s$ at asymptotic infinity, then this note predicts that the tension of the IIA/IIB wall \textit{vanishes} when $g_s=0$ for all $X^+$ as there is no lightlike dilaton profile to speak of which was the only contribution to the tension.

Other interesting avenues to pursue are understanding the IIA/IIB wall's worldvolume theory more systematically. Our construction suggests that the theory is non-local (due to the appearance of strings with vanishing Einstein frame tension), which could end up being necessary due to the fact that the chiral IIB fields have a mass generated on the IIA side; a symmetric mass generation feature that typically only happens at strong coupling for local field theories.

From more bottom-up considerations, typically constraints on the tension of branes predicted by the Swampland Cobordism Conjecture and/or charge completeness regard static branes rather than lightlike ones. Not to mention, given the subtitles for defining the tension of codimension-1 branes, statements such as the Elementary Constituents Conjecture \cite{Nevoa:2025xiq} do not yet take into account such scenarios. It would also be interesting to understand how this fits into more general bottom-up conditions on domain wall/cobordism defects connecting different quantum gravity vacua.

\section*{Acknowledgments}
I thank Markus Dierigl, Shota Komatsu, Miguel Montero, and Ashoke Sen for helpful discussions. I thank Jonathan J. Heckman, Jake McNamara, and Julio Parra-Martinez for related collaboration. The work of ET is supported in part by the European Research Council Starting Grant QGuide-101042568 - StG 2021.

\appendix
\section{IIB Worldsheet as an Orbifold of the IIA Worldsheet}\label{app:TheOrb}

In this Appendix, we review the classic fact that the IIB worldsheet CFT can be seen as an orbifold of the IIA worldsheet CFT by $(-1)^{F_L}$ and vice-versa. We work here in the lightcone Green-Schwarz (GS) string for relevance to the main text but these steps can equally be done in the lightcone and covariant RNS setting. As mentioned in the main text, the IIA GS fields are given by $X^i$, $\theta^{\Dot{\alpha}}_+$, and $\theta^{\alpha}_-$ which are respectively eight free scalars and eight free left-moving and right-moving Majorana-Weyl fermions. In the GS worldsheet, all fields are periodic along 1-cycles. So for a closed string with spatial coordinate $\sigma\sim \sigma+2\pi$ we have
\begin{equation}
  X^i(\sigma+2\pi)=X^i(\sigma) \quad \quad \theta^{\Dot{\alpha}}_+(\sigma+2\pi)=\theta^{\Dot{\alpha}}_+(\sigma), \quad \quad
  \theta^\alpha_-(\sigma+2\pi)=\theta^\alpha_-(\sigma)
\end{equation}
which in particular implies that the 2D fermions have zero-modes which satisfy
\begin{equation}
 \{ \theta^{\Dot{\alpha}}_{+,0}, \theta^{\Dot{\beta}}_{+,0}\}=\delta^{\Dot{\alpha} \Dot{\beta}}, \quad \quad
\{ \theta^{\alpha}_{-,0}, \theta^{\beta}_{-,0}\}=\delta^{\alpha \beta}
\end{equation}
The left-moving ground states can be obtained by acting on the left-moving vacuum $\ket{0}_L$ by $\theta^{\Dot{\alpha}}_{+,0}$ which fills out a $\mathbf{8}_v\oplus \mathbf{8}_s$ representation. Similarly, the right-moving ground states are in the $\mathbf{8}_v\oplus \mathbf{8}_c$ and the closed string ground states result from the tensor product of the left- and right-moving sectors. We see then that both the RR and NSNS sector states both appear from a single sector in the GS string.

Gauging by $(-1)^{F_L}_{IIA}$ then projects out the $\mathbf{8}_s\otimes \mathbf{8}_v$ and $\mathbf{8}_s\otimes \mathbf{8}_c$ ground states and their associated excited states. In the twisted sector, the left-moving Majorana-Weyl fermion satisfies
\begin{equation}
    \theta^{\Dot{\alpha}}_+(\sigma+2\pi)=-\theta^{\Dot{\alpha}}_+(\sigma)
\end{equation}
which means that the modes $\theta^{\Dot{\alpha}}_{+,r}$ are half-integer graded, $r\in \mathbb{Z}+1/2$.
Casimir energy of this sector in $\ell_s=1$ units is non-zero
\begin{equation}
    E_{\mathrm{Casimir}}=+8\left(- \frac{1}{24}\right)-8\left( \frac{1}{48}\right)=-\frac{1}{2}
\end{equation}
but there is no tachyon in the spectrum after enforcing the level-matching condition $N_L-1/2=N_R$. The twisted sector ground state is $\theta^{\Dot{\alpha}}_{+,-1/2}\ket{0}_L$ tensored with the right-moving $\mathbf{8}_v\oplus \mathbf{8}_s$ ground states. These states satisfy $N_L-1/2=N_R=0$ and are therefore massless. We have therefore reproduced the left-moving Ramond sector of the IIB string. Notice that one can straightforwardly repeat this procedure in the opposite direction by gauging by $(-1)^{F_L}_{IIB}$. This orbifolding straightforwardly generalizes to the 2D $\mathcal{N}=(8,8)$ symmetric orbifold SCFT, $\mathrm{Sym}^N(\mathbb{R}^8)$, where the $(-1)^{F_L}_{IIA}$ action relevant to us is simply the diagonal $(-1)^{F_L}_{IIA}$ action on $N$ copies of the GS lightcone CFT. Note that the level-matching condition is enforced in the MST picture because $L_0-\widetilde{L}_0\neq 0$ states gain an infinite mass in the $g_s\rightarrow 0$ limit \cite{Dijkgraaf:1997vv}.

\begin{figure}[t]
    \centering
    \includegraphics[scale=0.4]{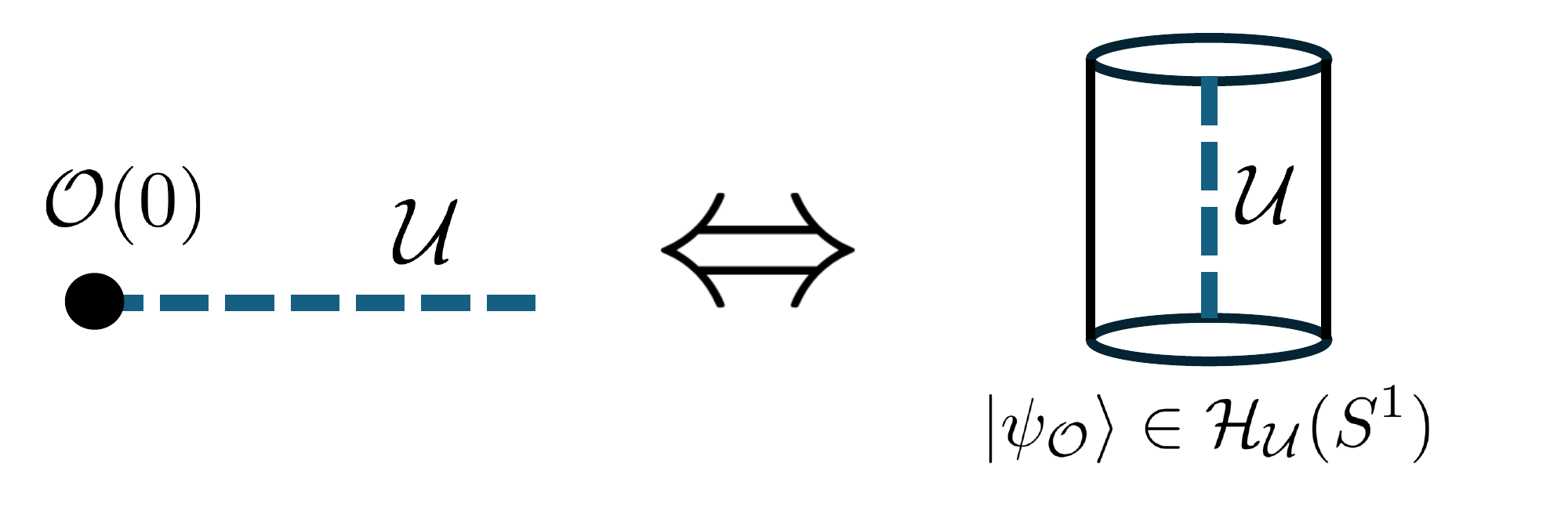}
    \caption{2D State-operator correspondence between a non-genuine local operator with a symmetry operator $\mathcal{U}$ attached and a state in a (topological) defect Hilbert space.}
    \label{fig:stateop}
\end{figure}

Before closing this appendix, we note that when gauging the IIA GS lightcone CFT by $(-1)^{F_L}_{IIA}$, one could still see the imprint of the projected out left-moving IIA Ramond sector by looking at a defect Hilbert space. In the gauged theory, the Wilson line operator for the $(-1)^{F_L}_{IIA}$ is precisely the topological line operator which generates the quantum global symmetry, $(-1)^{F_L}_{IIB}$. This means that ending this topological line would be gauge non-invariant with respect to the $\mathbb{Z}_2$ gauge symmetry but one can alleviate this by including a local operator odd under $(-1)^{F_L}_{IIA}$ which is also gauge non-invariant on its own. One can employ the state-operator correspondence to show that each of these non-genuine CFT operators are dual to a state in the defect Hilbert space, see Figure \ref{fig:stateop}. While such states are not relevant for GS string perturbation theory, the defect Hilbert space is part of the data off the symmetric orbifold CFTs, and therefore also for the Type II MSTs at $g_s=0$. The fate of these states when $g_s>0$ is explored in Section \ref{ssec:crossing}.




\bibliographystyle{utphys}
\bibliography{iiabwallmatrix}

\end{document}